# Thermopower and Unconventional Nernst in the Predicted Type-II Weyl Semi-metal WTe$_2$

K. Gaurav Rana[1*], Fasil K. Dejene[1*], Neeraj Kumar[1*], Catherine R. Rajamathi[2], Kornelia Sklarek[1], Claudia Felser[2], and Stuart S. P. Parkin[⊥1]

[1]Max Planck Institute of Microstructure Physics, 06120 Halle (Saale), Germany.

[2]Max Planck Institute for Chemical Physics of Solids, 01187 Dresden, Germany.

* These authors contributed equally.

[⊥]email: stuart.parkin@mpi-halle.mpg.de

Abstract

WTe$_2$ is one of a series of recently discovered high mobility semimetals, some of whose properties are characteristic of topological Dirac or Weyl metals. One of its most interesting properties is the unsaturated giant magnetoresistance that it exhibits at low temperatures. An important question is the degree to which this property can be ascribed to a conventional semi-metallic model in which a highly compensated, high mobility metal exhibits large magnetoresistance. Here we show that the longitudinal thermopower (Seebeck effect) of semi-metallic WTe$_2$ exfoliated flakes exhibits periodic sign changes about zero with increasing magnetic field that indicates distinct electron and hole Landau levels and nearly fully compensated electron and hole carrier densities. However, inconsistent with a conventional semi-metallic picture, we find a rapid enhancement of the Nernst effect at low temperatures that is nonlinear in magnetic field, which is consistent with Weyl points in proximity to the Fermi energy. Hence, we demonstrate the role played by the Weyl character of WTe$_2$ in its transport properties.

WTe$_2$ is a semimetal that has been calculated to exhibit several Weyl points ~50 meV away from the Fermi energy associated with Weyl cones that are tilted in energy-momentum space, a characteristic of a type-II Weyl semimetal.[1] The Weyl nature of Fermions in type-I Weyl semimetals, such as TaAs, has been confirmed by the observation of Fermi arcs in angle resolved photoemission spectroscopy (ARPES)[2]. By contrast, whilst Fermi arcs have been observed in WTe$_2$, these are not regarded as necessarily being indicative of a Weyl semimetal[3]. Rather the experimental signatures of Weyl fermions in WTe$_2$ to date, are from electrical transport measurements such as a chiral anomaly induced negative magnetoresistance and Weyl orbits from Fermi arcs observed using Shubnikov-de Haas oscillations[4, 5]. In addition to its topological properties, WTe$_2$ also exhibits a large non-saturating magnetoresistance (MR) at low temperatures, which is attributed to a nearly complete electron-hole (e-h) charge compensation together with large charge carrier mobilities[1]. Here we explore the semi-metallic and topological properties of WTe$_2$ using thermopower measurement which is a more sensitive probe of the electronic band structure than electrical transport. The longitudinal thermopower (S) is directly related to the energy derivative of the electrical conductivity at the Fermi energy $E_F$ via Mott's relation when $k_B T \ll E_F$ ($k_B$ = Boltzmann's constant; $T$ = Temperature)[6]. This is likely to be valid for the measurements presented here at low temperatures. The sign of S is determined by the electron (e) and hole (h) carrier densities and their corresponding effective masses. The electron ($S_e$) and hole ($S_h$) thermopowers compete with each other so that the net longitudinal thermopower is $S = S_e - S_h$. For an ideal Weyl semimetal, one expects $S$ to be close to zero and the competition between the $S_e$ and $S_h$ contributions can lead to distinctive values of S derived from the respective electron and hole Landau levels (LLs). On the other hand, the electron and hole contributions to the Nernst (transverse thermopower) are additive giving rise therefore to a much larger Nernst (N) effect, which is linear in magnetic field. A deviation from this linearity and a saturating response at a critical field could point to the presence of additional contributions from the Weyl points as observed in NbP [7], TaAs, TaP [8] and $Cd_3As_2$ [9] single crystals. The appearance of a large Nernst effect has been predicted for the case of Weyl semimetals (WSMs) [10], and is even more enhanced for the case of type-II WSMs [11]. In addition, a recent report [12] demonstrates that Nernst effect measurements can detect the influence of type-II Weyl points that are as much as ~100 meV above the Fermi level.

Previous studies of magneto-thermopower are limited to $WTe_2$ single crystals over a narrow temperature range [13, 14]. Performing magneto-thermopower measurements is challenging for semimetals since S is small and it is easy to pick up a contribution from the larger Nernst effect [6]. It is well established that to obtain reliable measurements of *S*, an isothermal sample mount is needed [7]. Here, we present temperature and magnetic field dependent measurements of the longitudinal and transverse thermopowers for isothermally mounted flakes of $WTe_2$ with various thicknesses (< 110 nm), each much smaller than the mean free path. We highlight important differences between the Seebeck measurements performed here and those previously reported for bulk $WTe_2$ samples.

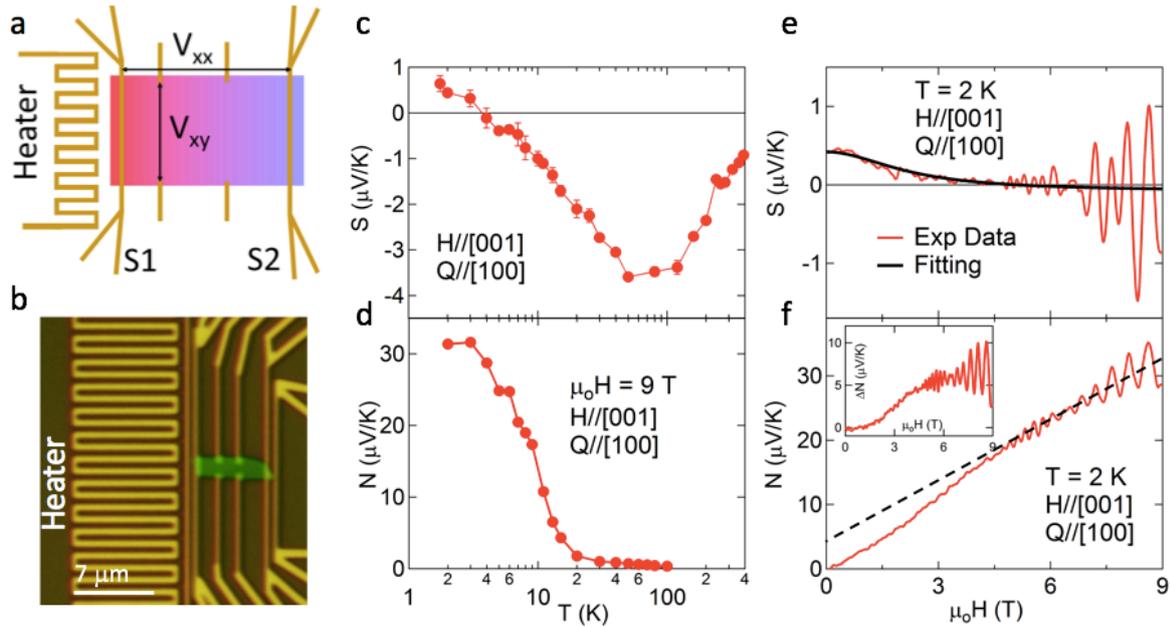

**Figure 1: Thermopower measurement and Nernst measurements:** (a) Schematic of the thermopower measurements. A temperature gradient $\nabla T \| x$ is applied along the length of the flake using a local heater, and an open circuit voltage is measured. Ru/Au sensors (S1 and S2) are employed to quantify the temperature difference between the hot and the cold ends of the flake. (b) Optical image of a typical device having a local heater and Au sensors and other electrical contacts. For magneto-thermopower measurements, a magnetic field is applied in the out of plane direction parallel to the c axis of the flake. (c) Seebeck coefficient (S) is plotted as a function of temperature for a 107 nm thick flake. (d) Temperature dependence of N at $\mu_0 H = 9$ T. (e) Magneto-Seebeck at 2 K. The black solid line shows a fit to Eq. 1. (f) Nernst coefficient measured as a function of field at 2 K. The black dashed line is a guide for the extrapolation of the high field Nernst value to B=0. The inset shows the non-linear part after subtracting the low field (< 3 T) linear slope.

Figure 1(a) shows a schematic of the $WTe_2$ device that was used to measure both magneto-thermoelectric and magneto-transport properties. The devices were prepared by first exfoliating a $WTe_2$ crystal and then transferring the (001) oriented exfoliated flakes onto a Si (001) wafer with a ~280 nm thick thermally grown $SiO_2$ layer, which was pre-patterned with alignment markers. These were later used to locate the flake so as to fabricate a resistive heater next to the flake as well as two temperature sensors, S1 and S2, on the "hot" and "cold" sides of the flake, and electrical contacts to the heater, sensor and the flake. All of these components of the device were fabricated from Au films that were deposited on Ru seed layers. The overall size of the device is a few microns in extent (see Methods for details). Note that the flakes were not encapsulated so that they are exposed to the environment[15]. To measure *S* and *N*, a temperature gradient $\nabla T$ is generated across the flake by applying a current to the heater and measuring the longitudinal ($V_{xx}$) and transverse ($V_{xy}$) open circuit voltages, as shown in the schematic diagram in Fig. 1 (a). The temperature gradient across the flake is then measured using S1 and S2. For the magneto - thermopower and -transport measurements, a magnetic field is applied in the out of plane direction along the flake's c axis. An optical image of a typical device is shown in Fig. 1(b).

Several devices were fabricated and studied. The thicknesses of the flakes varied from ~26 to ~107 nm, as measured using atomic force microscopy (AFM). The properties varied significantly but systematically with the thickness of the flake. We present the thickest flake (we will refer to this as device D1 which has a thickness of ~107 nm). The temperature dependence of $S$ is shown in Fig. 1 (c). Over a wide temperature range from ~4 to 400 K, we find that $S$ is small and negative which is indicative of electron carrier-dominated transport but a broad minimum near ~100 K suggests additional hole carrier transport. Indeed, at low temperatures below ~ 3 K, $S$ changes sign and becomes positive. We rule out its association with phonon drag as we estimate that any phonon drag contribution[16] would take place at much higher temperatures than 3 K (see Supporting Information, Section 5 for more details). A change in sign and a small magnitude of $S$ over the entire temperature range indicates comparable carrier densities of electrons and holes that are characteristic of a compensated semimetal. The temperature dependence of $S$ matches well with previous reports on bulk single crystals. Figure 1 (e) shows the out of plane magnetic field dependence of $S$ at 2 K. The most pronounced feature is the very large oscillations in $S$ as the field is increased above ~ 5 T. We find that $S$ oscillates around a value of approximately zero, changing its sign between successive maxima and minima. These oscillations, which have periods that vary from 81 to 151 T, as discussed later, clearly reflect field-induced quantization of the energy bands (Landau levels, LLs). The sign reversal of $S$ indicates the periodic dominance of respective electron and hole pockets as the field is increased. We note that, previously, bulk single crystals of $WTe_2$ were reported to have a very large magneto-Seebeck[13, 14] in contrast to our results. The earlier measurements on bulk samples were rather carried out in an adiabatic measurement configuration where a transverse temperature gradient arises due to the Righi-Leduc effect. This unwanted thermal gradient causes an additional voltage along the longitudinal direction due to the large Nernst effect in semimetals.[17] Because the Righi-Leduc effect and the Nernst effect are antisymmetric in $H$, the resulting induced Seebeck effect is symmetric in $H$ and, thus, this contribution cannot be easily disentangled from the actual S. Hence, in the case of semimetals where nearly compensated carriers are prevalent, thermopower measurements must be performed in an isothermal measurement configuration [7] which ensures a zero transverse thermal gradient as achieved in our measurements by anchoring the WTe$_2$ flakes to the substrate.

We find that the magneto-Seebeck data can be described at a given temperature by the following empirical expression derived from Liang et al.[9, 18],

$$S_{xx}(H) = S_o \frac{1}{1+\mu_e\mu_h B^2} + S_\infty \frac{\mu_e\mu_h B^2}{1+\mu_e\mu_h B^2} \quad \text{---(1)}$$

The data in Fig. 1(d) can be fit with a mean mobility, $\sqrt{\mu_e \mu_h} = 0.45$ m$^2$/Vs , and $S_o = 0.42$ $\mu V/K$ and $S_\infty = 0.05$ $\mu V/K$ , where $S_o$ and $S_\infty$ are the Seebeck coefficients at zero and very large magnetic field, respectively. The obtained mean mobility is in close agreement with mobility values found from magneto-transport data, as shown later.

The temperature dependence of $N$ at 9 T is shown in Fig. 1 (d) for device D1. $N$ was found from the measured transverse voltage ($V_{xy}$) using the formula, $= V_{xy} \frac{l}{w\Delta T}$, where $w$ and $l$ are the width and length of the flake, respectively (see Fig. 1a). $N$ takes very small values at higher temperatures but increases rapidly as the temperature is decreased below ~ 30 K (see Fig. 1(d)). At 2 K, $N$ is more than an order of magnitude larger (~ 30 times) than $S$.

The total Nernst coefficient per unit field ($\nu = N/H$), for negligibly small Hall conductivities, can be expressed as[19, 20]:

$$\nu = \frac{\sigma_{xx}^e \nu_e + \sigma_{xx}^h \nu_h}{\sigma_{xx}^e + \sigma_{xx}^h} + \frac{(S_e+S_h)(\mu_e+\mu_h)\sigma_{xx}^e \sigma_{xx}^h}{(\sigma_{xx}^e+\sigma_{xx}^h)^2} , \quad \text{--- (2)}$$

where $\nu_{e(h)}$, $S_{e(h)}$, $\mu_{e(h)}$ and $\sigma_{xx}^{e\,(h)}$ are the corresponding electron (hole) Nernst, Seebeck, mobilities and longitudinal conductivities, respectively. The first term arises from the individual bands, which is usually very small[17], and, the second term is an ambipolar term that originates from the presence of both e and h bands. This term is usually large for semimetals. For a perfectly compensated system, and assuming $\mu_h \equiv \mu_e$, the second term will dominate.[17] For a simple case when the conductivities depend linearly on energy[21] and using $S_e = \pi^2 k_B^2 T/3e\epsilon_F$, we obtain:

$$\frac{N}{H} = \nu = \frac{\pi^2}{3}\frac{k_B^2 T}{e}\frac{\mu_e}{\epsilon_F} \text{ --- (3)}$$

Thus, a large Nernst response is found in systems with large carrier mobilities and small Fermi energies as in WTe$_2$ at low temperatures[1]. From equation (3) *N* should vary linearly with *H*. Clear quantum oscillations in *N* are visible above 4 T superposed on a non-linear Nernst signal (see Fig. 1(f)). In addition, a clear deviation from a linear scaling of *N* with *H* as shown in the inset of Fig. 1 (f) that is at odds with Eq. (3). We also note that equation (3) is similar to that used in Ref [9], using only a single band model that showed a universal scaling between $\nu$ and $\frac{\mu}{\epsilon_F}$ for different materials.

In order to gain more understanding of the S and N data presented above, we derive the carrier mobilities from magneto-transport measurements. The magnetoresistance (MR) measured at various temperatures for D1 is shown in Fig. 2(a), where MR is defined as MR % $= 100 \frac{\rho_{xx}(H)-\rho_{xx}(0)}{\rho_{xx}(0)}$ and $\rho_{xx}(H)$ is the resistivity. The flakes show a large quadratic variation of resistance on field although with deviations in small fields and clear quantum oscillations at high fields. These oscillations indicate high carrier mobilities in these flakes[13,14].

The maximum value of MR is ~1450 % at 2 K in the maximum field available of 9 T. We find that the MR is reduced systematically as the flake thickness is reduced. For the thinnest flake (~26 nm) that we measured, the MR was ~ 140% in 9T at 2 K. These MR values are much lower than those found in bulk single crystals[22-25] but are comparable to those previously reported in un-encapsulated WTe2 flakes of comparable thickness[26-28]. Very recent studies suggest that encapsulation of the top and bottom of the flakes with *h*-BN gives larger MR but still much smaller than those found for single crystals themselves[29].

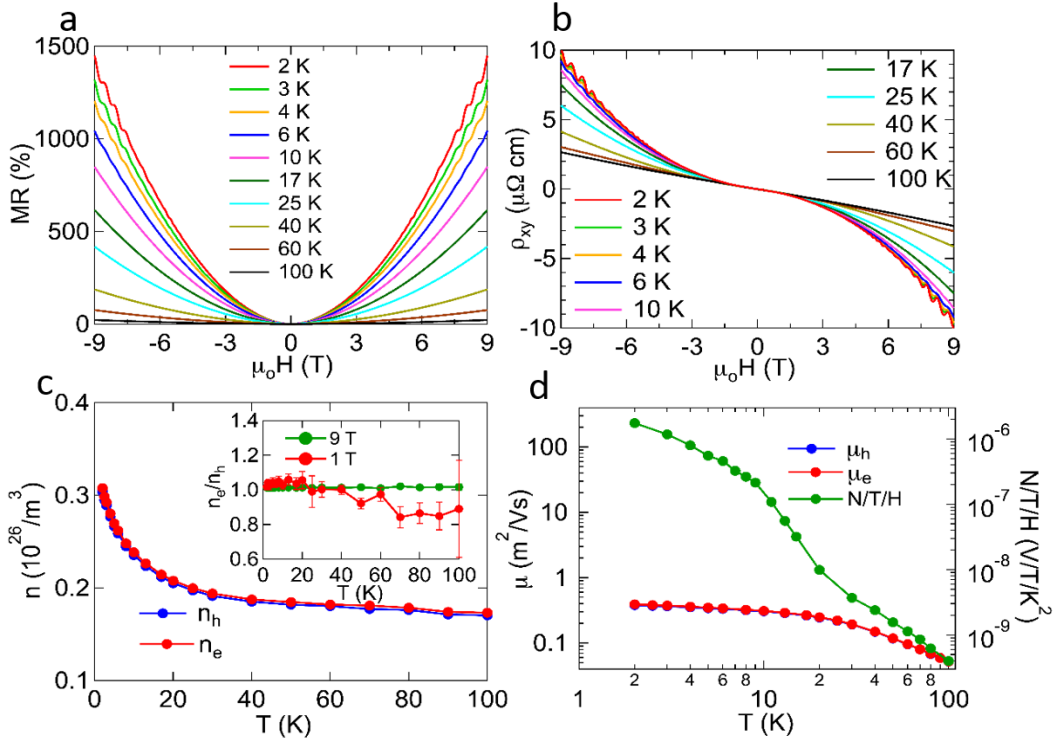

**Figure 2: Magneto-transport and Hall measurements:** (a) Resistivity measured as a function of magnetic field (H ∥ *c* axis) for a 107 nm thick flake. (b) Hall resistance measured at various temperatures as a function of field. (c - d) Using Eqn. 4 and 5, $\rho_{xx}$ and $\rho_{xy}$ are simultaneously fit to obtain the carrier concentrations ($n_e, n_h$) and mobilities ($\mu_e, \mu_h$) at various temperatures. N/T/H is included in (d).

The Hall resistivity $\rho_{xy}$ for device D1 is shown in Fig. 2(b) for temperatures up to 100 K. Similar to the MR quantum oscillations are observed in the Hall data. The frequencies of the oscillations are similar for both sets of data, indicating a common origin. At 100 K, $\rho_{xy}$ varies almost linearly with H but as T is reduced there is a strong non-linear component that increases with decreasing temperature. The non-linearity in $\rho_{xy}$ is consistent with multiple carrier transport. Therefore, we consider the simplest semi-classical model with one electron-like and one hole-like band. In this two-band model $\rho_{xx}(H)$ and $\rho_{xy}(H)$ are described by the following equations at a given temperature.

$$\rho_{xx}(H) = \frac{1}{e}\frac{(n_e\mu_e+n_h\mu_h)+(n_h\mu_e+n_e\mu_h)\mu_e\mu_h H^2}{(n_e\mu_e+n_h\mu_h)^2+[(n_h-n_e)\mu_e\mu_h]^2 H^2} \qquad (4)$$

$$\rho_{xy}(H) = \frac{H}{e}\frac{(n_h\mu_h^2-n_e\mu_e^2)+(n_h-n_e)\mu_e^2\mu_h^2 H^2}{(n_e\mu_e+n_h\mu_h)^2+[(n_h-n_e)\mu_e\mu_h]^2 H^2} \qquad (5)$$

We simultaneously fit the non-oscillatory part of $\rho_{xx}$ and $\rho_{xy}$ data to find the carrier concentrations ($n_e, n_h$) and mobilities ($\mu_e, \mu_h$) at various temperatures, as shown in Fig. 2 (c). (See Supporting Information, Section 3 for details of the fitting procedure used). The mobilities and carrier densities can be readily found when the mobilities are high enough that the quadratic terms in $H^2$ in Eq. 4 and Eq. 5 take a significant value. This is the case for temperatures below ~100 K. At 2 K we find that the electron and hole carrier concentrations are nearly equal with a slight preponderance of electrons. The slight excess in $n_e$ is clearly indicated by the negative coefficient of the $H^3$ term in high fields. With regard to the mobilities, we find that $\mu_e$ is close to $\mu_h$ for all temperatures below 100 K with $\mu_e$ being about 2% bigger than $\mu_h$. The error bars in $\mu_e$ and $\mu_h$ are estimated to be ~2.5%.

Although the two-band model can fairly well describe the magneto-transport data, nevertheless a much better fit is obtained for fields below 1 T. First, we performed the two-band model fitting in two field ranges of up to 1 T and up to 9 T to thereby obtain $n_e, n_h, \mu_e$ and $\mu_h$ (at zero field) at each temperature (Fig. S4). The overall trend in the temperature dependence of each of them is the same for both fitting procedures but the fitting performed over the larger field range seems to overestimate the carrier densities and underestimate the carrier mobilities as compared to fits that use the smaller field range: hence, the ratio $\frac{n_e}{n_h}$ is not the same for each of them. By comparing fits for several ranges at each temperature, we observe that there is a possibility that, at high fields, some extra energy band might have been picked up causing either a change in mobilities attested to by the deviation from the two-band behavior, and/or changes in carriers, for instance, due to field induced changes in the Fermi surface. Empirically, the MR is better described by a sub-quadratic ($\propto H^{1.9}$) power law dependence above 1 T, as found earlier for WTe$_2$ [24, 28, 29]. A similar sub-quadratic power law has been found for other non-saturating large MR materials such as PtSn$_4$[30], NbSb$_2$[31], WP$_2$[32] and is observed in graphite, where it is ascribed to an impurity insensitive scattering mechanism.[33]

We note that $\mu_e$, $\mu_h$ (derived from MR) and N/T both increase as the temperature is reduced, although the increase in $\frac{N}{T}$ is much more rapid below ~ 30 K, as shown in Fig. 2(d). Usually, such an enhancement could be attributed to improved charge compensation or larger relaxation times (carrier mobility). Nevertheless, the rapid enhancement of the Nernst signal is at odds with the relatively slower mobility increase and, hence, calls for an additional Nernst contribution. The appearance of a large Nernst effect has been predicted for the case of Weyl semimetals [10] [11]. Similarly, as a function of magnetic field, the non-linear Nernst saturates at fields larger than 8T (inset of Fig. 1(f)).

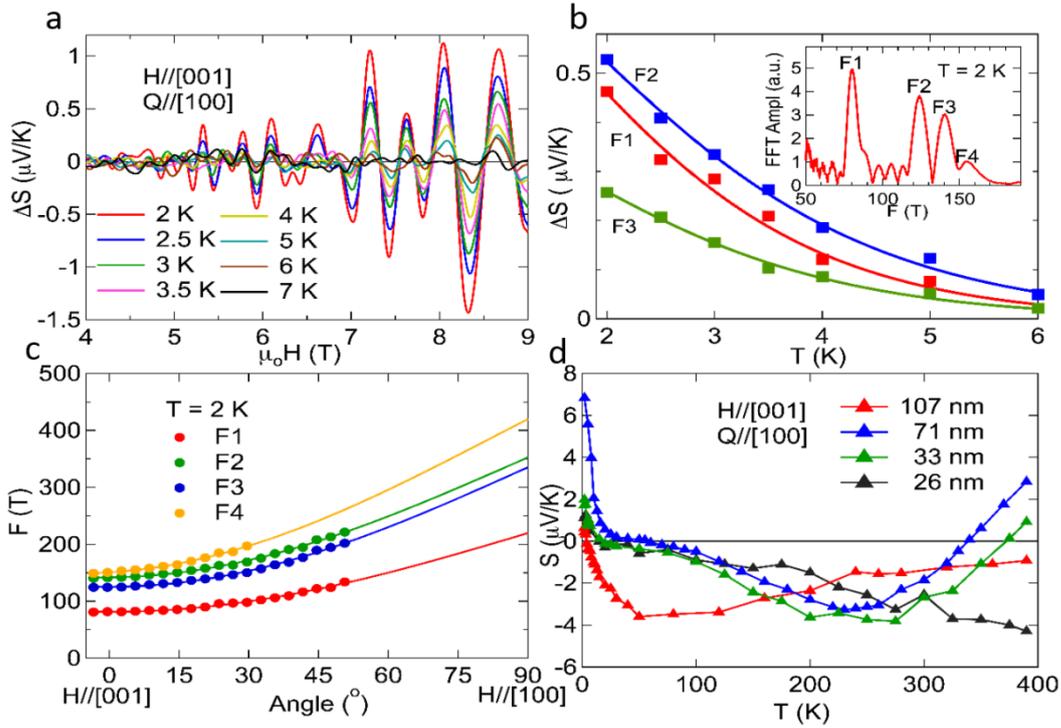

**Figure 3: Temperature and thickness dependence of quantum oscillations in the Seebeck coefficient.** (a) Oscillatory part of the Seebeck coefficient for a 107 nm thick flake as a function of field for various temperatures. (b) Amplitudes of Fast Fourier transform (FFT) derived oscillations versus temperature. A typical FFT is shown in the inset at 2K. (c) Angle dependence of the FFT derived frequencies. The solid lines are a sinusoidal fit. (d) S versus temperature for several devices with difference flake thicknesses.

By extrapolating this Nernst signal to B = 0 (as shown with a dashed line in Fig. 1(f)), we obtain a value $N_0 = 4.2$ μVK$^{-1}$, which emanates from the non-zero Berry flux of the Weyl points in WTe$_2$. This observation further confirms the Weyl physics in WTe$_2$. A similar but more pronounced enhancement has been recently observed in the Dirac semimetal, Cd3As2 (below 50 K)[9] and the Weyl semimetal NbP (near 100 K)[7]. In both these cases, a nonlinear dependence of N on magnetic field is also observed, as we also find in WTe$_2$. In Cd$_3$As$_2$ and NbP, these features have been shown to be directly associated with the Berry curvature of the Weyl points and chemical potential shift induced Dirac dispersion proximity, respectively[7, 9]. Recent measurements of the Nernst effect in TaP and TaAs [8] also confirm that the characteristic saturation plateau in Nernst voltage beyond a critical magnetic field is a direct consequence of a finite Berry curvature and is indicative of Weyl points proximal to E$_F$ at low temperatures, similar to what we observe for WTe$_2$ at low temperatures.

In Fig 3a, we show the oscillatory component of S. From Fast Fourier Transforms (FFT) we identify four distinct frequencies: 82 T, 124 T, 141 T, and 151 T. The first and fourth frequencies correspond to electron pockets and the other two to hole pockets. Similar frequencies (See Supporting Information, Table 1) are observed from the three other transport coefficients showing good agreement between these various properties, as expected.

We explore the shape of the Fermi surface by measuring quantum oscillations in S while rotating the magnetic field vector from H ∥ c to H ∥ a (see Supporting Information, Section 2). Each of the oscillation frequencies, which are proportional to an extremum cross-section of the Fermi surface, increases as the field becomes parallel to the a-axis, indicating an anisotropic Fermi surface. For an ellipsoidal Fermi surface, a sinusoidal angle dependence is expected (solid lines in Fig. 3(d)), as also reported by Zhu *et al* in bulk WTe$_2$[5]. We estimate the volume of the Fermi surface, and, hence, the carrier density from the average frequency. The carrier densities are thereby calculated to be $0.74 \times 10^{20}$ cm$^{-3}$ and $0.66 \times 10^{20}$ cm$^{-3}$, for electrons and holes respectively. These are of the same order of magnitude as the corresponding values of $0.329 \times 10^{20}$ cm$^{-3}$ for electrons and $0.325 \times 10^{20}$ cm$^{-3}$ for holes, obtained from the simultaneous fit of the magneto-transport and Hall effect data within a semiclassical two carrier model. Note that the ellipsoidal Fermi pocket approximation gives an upper bound to the carrier densities. From the amplitude of the quantum oscillations in S as a function of temperature which disappear above 7 K, as shown in Fig. 3(b), we estimate the effective mass of the electron and hole carriers using the Lifshitz-Kosevich formula[34] to be between 0.49 - 0.57 $m_e$, (where $m_e$ is the free election mass) with a corresponding Fermi energy of ~17 to 32 meV.

Fig. 3 (d) shows the thickness dependence of S. The overall behavior of S as a function of temperature remains the same irrespective of the thickness, except that the broad minima in S shifts to lower temperatures for the thicker samples. The oscillations in Seebeck around zero have been also observed for other thicknesses.

We conclude that the magneto-thermal and magneto-transport properties of WTe$_2$ flakes of various thicknesses can be self-consistently accounted for within a semi-metallic model for which the electron and hole carrier densities and mobilities are very similar over a wide temperature range. However, at temperatures below ~30 K large deviations of the Nernst effect are found which are indicative of the Weyl nature of WTe$_2$. Most importantly, the nonlinear Nernst response, which has been theoretically predicted in Weyl semi metals and observed in type-I Weyl systems, is further evidence for this conclusion. Our results highlight the relevance of Nernst and Seebeck measurements to capture features of recently discovered high mobility semimetals which could otherwise not be visible in other transport properties.

**Methods:**

High-quality single crystals of WTe$_2$ are grown from a Tellurium flux using a chemical vapor transport method. The quality of the as-grown single crystals was confirmed using X-Ray diffraction (XRD) and scanning electron microscope (SEM). Thin WTe$_2$ flakes were exfoliated from a small single crystal onto a pre-patterned Si/SiO$_2$ substrate in a clean room environment. Within a few seconds after exfoliation, the samples were spin-coated using a ZEP-520 resist to minimize contact with air. After patterning the devices using electron beam lithography, 10 nm Ru/ 100 nm Au contacts were deposited. A uniform temperature gradient is applied in the *a-b* plane of the flake using a meander heater. By measuring the longitudinal and transverse open circuit voltages, and by carefully calibrating two pairs of Ru/Au temperature sensors to obtain the temperature gradient, the Seebeck and Nernst coefficients are determined. The substrate on which the flakes lie acts as a heat sink to ensure an isothermal measurement configuration. All the measurements are performed using Keithley 2450 source meters and Keithley 2182A nanovoltmeters in a commercially available Quantum Design Dynacool cryostat in vacuum.

**Supporting Information Available**

Supporting Information contains details about the experimental procedures for the device fabrication and measurements. The angle dependence of quantum oscillations, analysis of two carrier model, Kohler's plot and calculation of Debye temperature are also discussed.

**Acknowledgement:**

We acknowledge partial support from the ERC Advanced Grant No. 670166 "SORBET" and the ERC Advanced Grant No. 742068 "TOPMAT".

**Conflict of Interests:** The authors declare no competing financial interest.